\documentclass{elsart}
\usepackage{epsfig}
\def\ltsima{$\; \buildrel < \over \sim \;$}
\def\simlt{\lower.5ex\hbox{\ltsima}}
\def\gtsima{$\; \buildrel > \over \sim \;$}
\def\simgt{\lower.5ex\hbox{\gtsima}}
\def\ls{{_<\atop^{\sim}}}

\def\cgs{ ${\rm erg~cm}^{-2}~{\rm s}^{-1}$ } 
\begin{document}
\runauthor{F. Fiore et al.}
\begin{frontmatter}
\title{Spectroscopic identification of ten faint 
hard X-ray sources discovered by {\it Chandra}
\thanksref{thankeso}}
\thanks[thankeso]{Based on observations 
carried out at the ESO--La Silla Telescopes}
\author[oar,sdc,cfa]{F. Fiore}
\author[roma3]{F. La Franca}
\author[oab,unibo]{C. Vignali} 
\author[oab]{A. Comastri}
\author[roma3]{G. Matt}
\author[roma3]{G.C. Perola}
\author[tesre,cfa]{M. Cappi}
\author[cfa]{M. Elvis}
\author[oar,cfa]{F. Nicastro}
\address[oar]{
   Osservatorio Astronomico di Roma, Via Frascati 33,
        I--00044 Monteporzio Catone, Italy}
\address[sdc]{BeppoSAX Science Data Center, Via Corcolle 19, I--00131 
Roma, Italy}
\address[cfa]{Harvard-Smithsonian Center for Astrophysics, 
60 Garden st., Cambridge MA 02138 USA}
\address[roma3]{Dipartimento di Fisica, Universit\`a degli Studi ``Roma Tre",
Via della Vasca Navale 84, I--00146 Roma, Italy}
\address[oab]{Osservatorio Astronomico di Bologna, Via Ranzani 1,
I--40127 Bologna, Italy}
\address[unibo]{Dipartimento di Astronomia, Universit\`a di Bologna, 
via Ranzani 1, I--40127 Bologna, Italy}
\address[tesre]{Istituto Tecnologie e Studio Radiazioni Extraterrestri, 
CNR, Via Gobetti 101, I--40129 Bologna, Italy}

\begin{abstract} 
We report optical spectroscopic identifications of 10 hard (2--10 keV)
X-ray selected sources discovered by {\it Chandra}. The X-ray flux of
the sources ranges between 1.5 and 25 $\times10^{-14}$ \cgs, 
the lower value being 3 times fainter than in previous BeppoSAX and 
ASCA surveys.  Their R band magnitudes are in the range 12.8--22.  
Six of the {\it Chandra} sources are broad line quasars with 
redshifts between 0.42 and 1.19, while the optical identification of 
the remaining four is quite varied: two are X-ray obscured, emission 
line AGN at $z$=0.272 and $z$=0.683, one is a starburst galaxy at $z$=0.016 
and one, most unusually, is an apparently normal galaxy at $z$=0.158.  These
findings confirm and extend down to fainter X--ray fluxes the BeppoSAX
results, in providing samples with a wide range of X-ray and optical
properties.

The ratio between the soft X-ray and the optical luminosity of the $z$=0.158
galaxy is a factor at least 30 higher than that
of normal galaxies, and similar to those of AGN.  The high X--ray
luminosity and the lack of optical emission lines suggest an AGN in
which either continuum beaming dominates, or emission lines are
obscured or not efficiently produced. 
\end{abstract}


\end{frontmatter}

\section{Introduction}

The {\it Chandra} X-ray Observatory was launched on July 23 1999,
carrying on board a revolutionary high resolution mirror assembly,
with a Point Spread Function of 0.5 arcsec (half power radius) over
the broad 0.1 to 10 keV band (Van Speybroeck, et al. 1997).  This,
together with the aspect camera which at the moment provides attitude
solutions with errors of the order of 1-2 arcsec\footnote{When the
data are definitively reprocessed, the aspect for image reconstruction
should be $\sim0.5$ arcsec and the source positions should be better
than 1 arcsec}, allows the study of spatial extent of X-ray sources on
similar scales, i.e. smaller or similar to the size of a L$^*$ galaxy
at any redshift; and gives X-ray source positions at least as good as
2--3 arcsec, immediately allowing the unambiguous identification of the
optical counterparts of faint X-ray sources.  Consequently, the
determination of the source redshifts via optical spectroscopy becomes
highly efficient.

The improvement provided by {\it Chandra} is especially significant in
the hard (2--10 keV) X-ray band.  Surveys of the hard X-ray sky have
been performed in the past by ASCA and BeppoSAX (Ueda et al. 1998,
Della Ceca et al. 1999, Fiore et al. 2000a, Giommi et al. 2000,
Comastri et al. 2000).  However, the large error boxes (1-2 arcmin)
limited the optical identification process to classes of objects with
low surface density, at a given optical magnitude limit. As a result,
most of the identified sources are emission line AGN (Fiore et
al. 1999, Akiyama et al. 2000, and La Franca et al. in
preparation). About 30 \% of the BeppoSAX HELLAS survey sources studied
spectroscopically down to R=20.5 have escaped a secure identification
(Fiore et al. 2000b, La Franca et al. in preparation), although many
normal galaxies and stars have been observed in these error-boxes.

{\it Chandra}'s unprecedented capabilities make identifications
unambiguous, and open up the possibility of searching for and studying
classes of sources not previously recognized as strong hard X-ray
emitters, and of assessing their contribution to the hard X-ray cosmic
background (XRB; e.g. Griffiths \& Padovani 1990).  In particular, it
will be possible for the first time to begin studying normal galaxies
at $z>0.1$, as well as possible ``minority'' hard X-ray source
populations (Kim \& Elvis 1999).

We have started a pilot project of spectroscopic identification 
of {\it Chandra} sources in two medium--deep fields
 that were visible from La Silla in January 2000, with
the aim of verifying the feasibility of such studies with 4m class
telescopes.  The results on 10 X-ray sources are very encouraging and
are described in the following.

\section{X-ray data}

The {\it Chandra} X-ray Observatory consists of four pairs of
concentric Wolter I mirrors reflecting 0.1-10 keV X-rays into one of
the four focal plane detectors: ACIS-I, ACIS-S, HRC-I or HRC-S
(Weisskopf, O'Dell \& VanSpeybroeck 1996). 
All the results presented in the following were
obtained with the $16'\times16'$ ACIS-I CCD instrument.  Table 1 gives
the log of the Chandra observations. One field is centered on the
$z$=0.6 quasar PKS0312$-$77, the other is located $180^o$ away from the
radiant point of the Leonid meteor shower.
Level 2 processed data (Fabbiano et al. in preparation) were obtained
from the Chandra public archive.  Data were cleaned and analyzed using
the CIAO Software (release V1.1, Elvis et al. in preparation).  Time
intervals with large background rate were removed, as well as hot
pixels and bad columns. Only the standard event grades 
(0, 2, 3, 4 and 6) were used.

Sources were detected in images accumulated in the 2--10 keV band
(channels 136-680).  Robust sliding-cell algorithms were used to
locate the sources.  We used both the $celldetect$ program available
in the CXC data analysis package CIAO (Dobrzycki et al. 1999) and a
variation of the DETECT routine included in the XIMAGE package (Giommi
et al. 1991).  The method consists in first convolving the X-ray image
with a wavelet function, in order to smooth the image and increase
contrast, and then running a standard sliding cell detection method
on the smoothed image.  The quality of each detection was checked
interactively.

Final net counts are estimated from the original (un--smoothed) image,
to preserve Poisson statistics. A binning factor of 2 (i.e. pixels of
$\sim$ 1 arcsec) and source box sizes of 6 arcsec (offaxis angle
$<5'$), 10 arcsec ($ 5'<$ offaxis angle $<10'$) and 30 arcsec (offaxis
angle $>10'$) maximize the signal to noise ratio, given the local
background, PSF and source intensity.  The background is calculated
using source-free boxes near the sources. This is usually very low in
the 2--10 keV band, i.e. 0.15--0.3 counts per 6 arcsec detection 
box side, per 10 ks.

For this study only sources with more than 10, 15 and 40 counts in the
6 arcsec, 10 arcsec and 30 arcsec detection cells respectively, are
considered.  Given a conservative background of 0.6, 1.5 and 15 counts
in the detection cells this corresponds to a Poisson probability for a
background fluctuation of $10^{-9}$ in all cases. The total number of
detection cells can be estimated in about 7800 6$''$ cells plus 5000
10$''$ cells plus a few tens of 30$''$ cells in each observation (4
ACIS-I chips).  Therefore, the total number of spurious sources in
the two fields (0.14 deg$^2$) at the chosen detection thresholds is absolutely
negligible (10$^{-5}$). 
Thirteen sources detected in the 2--10 keV images passed
these thresholds. All but one are also detected in the softer 0.5--2 keV
band (the non--detected source has about 6 counts in the soft band).

Count rates were corrected for the PSF and the telescope vignetting
calculated at 4.5 keV (2--10 keV band) and 1.5 keV (0.5--2 keV band)
using Figures 3.8 and 3.10 of the AXAF Observatory Guide (1997).
The correction is about 25 \% at an off-axis angle of 12 arcmin 
and 10 \% at 6 arcmin.
Fluxes in the 2--10 keV band were computed assuming a count rate to
flux conversion factor of $2.5\times10^{-11}$ \cgs per count
s$^{-1}$. Fluxes in the 0.5--2 keV were computed assuming factors of
$5.2\times10^{-12}$ and $4.6\times10^{-12}$\cgs per count s$^{-1}$ for
Leonid Anti-Rad and the PKS0312$-$770 fields respectively.  These
conversion factors are appropriate for power law models with
$\alpha_E=0.7$ (2--10 keV) and $\alpha_E=1.0$ (0.5--2 keV) respectively,
assuming the Galactic absorbing column densities quoted in Table 1.
Table 2 gives for the thirteen detected sources the X-ray position and
off-axis angle, the 2--10 keV and 0.5--2 keV counts (corrected for the
PSF and the vignetting) and the corresponding fluxes.

\begin{table*}
\centering
\caption{Chandra observations}
\vspace{0.05in}
\footnotesize
\begin{tabular}{lcccc}
\hline
~& ~ & ~ & ~ & ~\\
Field & $N_H$(Gal) & Date  & ACIS & Exposure  \\
      & $10^{20}$ cm$^{-2}$ & & & seconds \\
\hline
PKS0312-770 & 8.0 & 1999 Sept. 8 & ACIS-I, 4 chips & 14789 \\
LEONID ANTI-RAD & 2.5 & 1999 Nov. 17 & ACIS-I, 4 chips & 20630\\
\hline

\end{tabular}

\end{table*}

\begin{figure*}
\centerline{
\epsfig{file=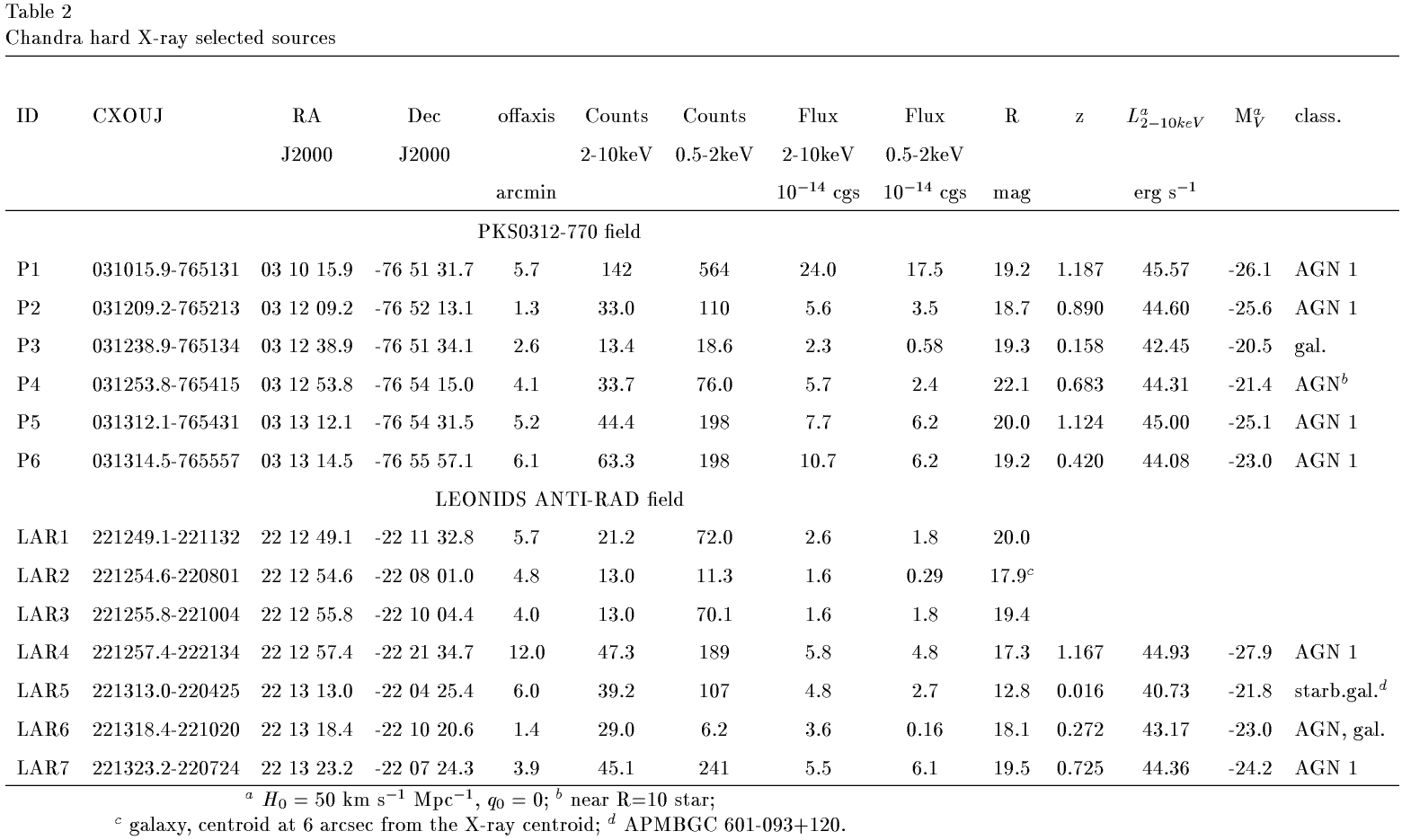, height=21cm,width=21cm, angle=90 }
}
\end{figure*}

\setcounter{figure}{0}

\section{Optical spectroscopy}

Optical counterparts were found in the USNO catalogue down to 
R$\sim 20$ in ten cases, in the DSS2 E(R) images in two cases 
(LAR1 and LAR3), and on EFOSC2 R band image in the last case (P4).  
Optical counterparts were found 
for 11 sources with a typical displacement of $\ls2$ arcsec.  In one
case (LAR4) the displacement is of 4''. We consider the identification
secure because a) LAR4 is observed by ACIS at an offaxis angle of 12
arcmin, where the PSF is highly asymmetric and broader than 10 arcsec,
making the X-ray position more uncertain, and b) the optical source is
a bright, broad line quasar, the surface density of which is so low
that the probability to find one by chance at $4''$ from an X-ray
source is negligible. In fact, the probability of finding {\it
any} R$\leq$22 galaxy in a $4''$ radius error-box by chance is $<$ 0.04,
while the same probability for an AGN is at least ten times smaller.
Therefore the number of AGN misidentifications in our sample is $<0.05$. 
The number of expected galaxy 
misidentifications at R$\leq$20 is $<$ 0.2, while that of R$\leq$22 is $<$ 0.5.
In one case (LAR2) no optical counterpart brighter than 
R $\sim$ 20 is present in the X--ray error box.
However a R=18 galaxy, with an optical extension of 
about 3 arcsec, is present at 6 arcsec from the X-ray position. 
Therefore the galaxy outskirts
touch the X-ray errorbox. We conservatively consider this identification 
uncertain.

We obtained long slit spectra of ten counterparts
\footnote{the three objects not observed in  the Leonid anti-rad field 
have optical magnitudes and X-ray fluxes similar to those of the ten
observed objects. The four sources observed in this field were
selected by chance}, using EFOSC2 at the ESO 3.6m telescope on January
3-6 2000.  Spectra were obtained using grism N.6 in the 3800-8100
$\AA$~range and a slit 2 arcsec wide, corresponding to a resolution of
26 \AA\ .
  
The complete set of flux-calibrated spectra is shown in Figure 1.  R
magnitude, redshift, optical and X-ray luminosity, and a
classification of the optical spectrum are given in Table 2.
Classification of narrow line objects is done using standard line
ratio diagnostics (e.g. Osterbrook 1981, Tresse et al. 1996), in
particular the [OIII]$\lambda5007$/H$\beta$,
[SII]$\lambda6725$/H$\alpha$ and
[OIII]$\lambda5007$/[OII]$\lambda3727$ line ratios.

\begin{figure*}
\centerline{
\epsfig{file=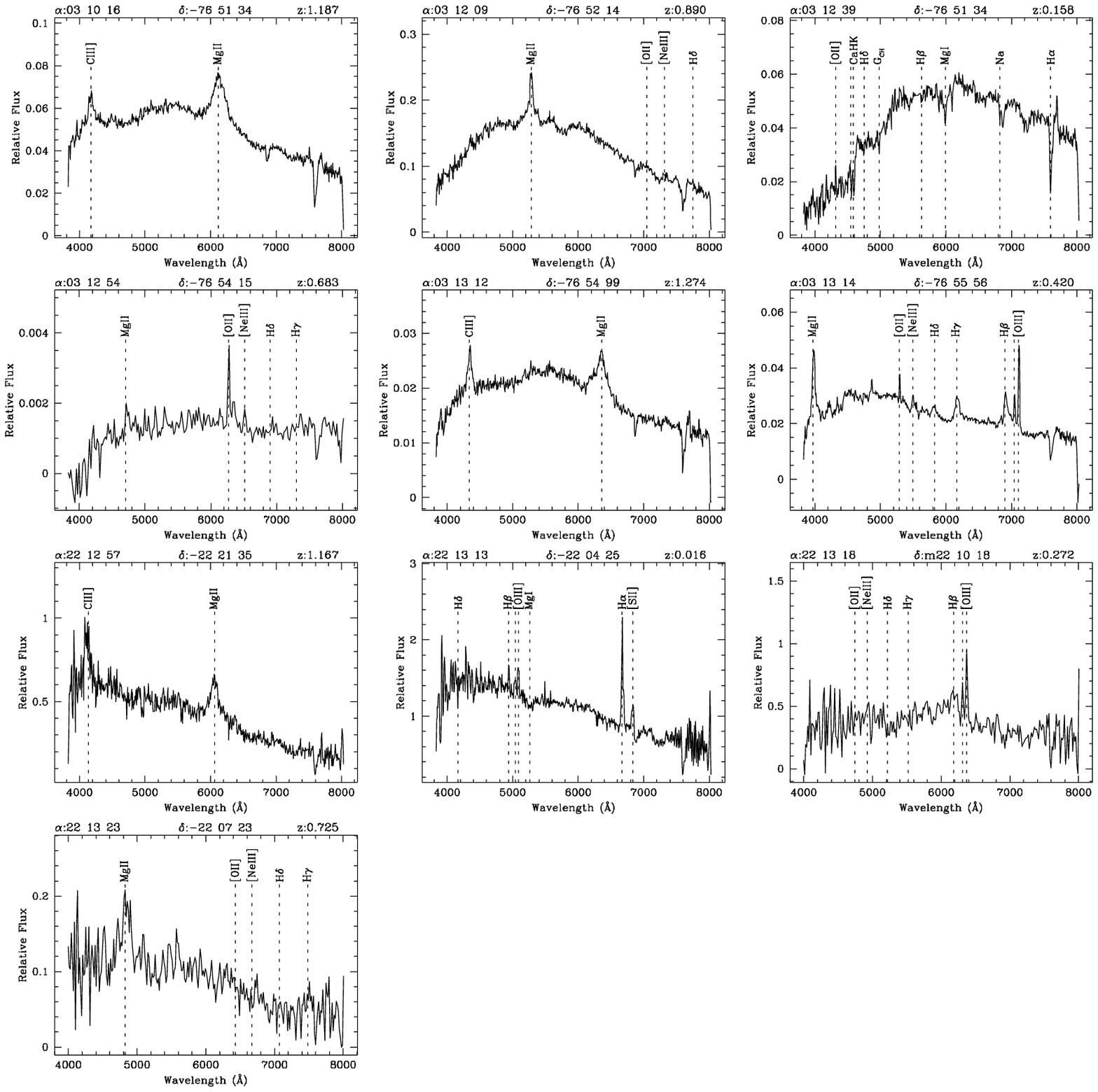, height=16cm,width=16cm, angle=0 }
}
\caption{
The EFOSC2 spectra of the ten spectroscopically identified 
{\it Chandra}  hard X-ray selected sources. Vertical dashed lines
are the most important expected atomic transitions, and are reported only 
for reference. 
}
\end{figure*}

We find six broad line quasars at redshift between 0.42 and 1.19 and
2--10 keV X--ray luminosities in the range $10^{44}-4 \times 10^{45}$ 
erg s$^{-1}$. The four remaining source identifications
are as follows:

LAR5 is identified with a bright $z$=0.016 starburst galaxy with an
X-ray luminosity of $\sim6\times10^{40}$ erg s$^{-1}$, similar to that
of other starburst galaxies of similar optical luminosity (Ptak et
al. 1999).

In LAR6, although [OIII] and $H\beta$ are detected, the spectrum is
rather noisy and it is not clear if $H\beta$ has broad wings.  Optical
classification is therefore uncertain.  The 2--10 keV luminosity of
$\sim2\times10^{43}$ erg s$^{-1}$ and the X-ray softness ratio (see
below) suggest that LAR6 is an obscured AGN.

P3 is identified with a normal galaxy without strong emission lines
(only a weak [OII] line with an equivalent width of $5\pm1$ \AA\ is apparent
in the spectrum).  The X-ray and optical luminosities are
$L_{2-10keV}\sim 3\times10^{42}$ erg s$^{-1}$ and $M_V=-20.5$. The
galaxy continuum is very red, suggesting an early type galaxy. The
Calcium break is 0.44$\pm0.11$, consistent with little or no dilution
from non--stellar light. A faint H$\delta$ is detected in absorption
(equivalent width of $-3.8\pm2.3$ \AA).

The spectrum of P4 is rather noisy because the source lies only 9 arcsec
from a 10 mag star, which strongly enhances the background.  We
identify P4 with an AGN at z=0.683 thanks to a rather strong [OII]
and weaker [NeIII] and MgII emission lines.
The MgII to [OII] intensity ratio of $\sim 0.9$ is 
much smaller than in broad line quasars and similar to those of type 2
AGN (Woltjer 1990). The 2--10 keV luminosity of $\sim2\times10^{44}$
erg s$^{-1}$ and the X-ray softness ratio (see below) suggest that P4
is a moderately obscured quasar.

We have searched the NVSS, IRAS faint source catalog, clusters of
galaxies (ACO, Abell, Corwin \& Orowin 1989) , 
normal galaxies, stars and AGN catalogs,
finding only one coincidence: LAR5 is identified with an APM galaxy.

\section {X-ray and optical properties}

\subsection{X-ray versus optical imaging}

The high spatial resolution of the {\it Chandra} telescope allows
for the first time a comparison of X-ray and optical images at similar,
arcsec, resolution.

We have studied the spatial extent of the X-ray sources by comparing
their count profiles with the {\it Chandra} PSF, taking into account
its dependence on the off-axis angle (as calibrated on the ground).  The
spatial extent is smaller than a few arcsec in all cases and therefore 
consistent with their being point sources.  Conversely,
at least three of the optical counterparts are extended galaxies
(``gal.'' in Table 2).

Figures 2a,b show the X-ray (2--10 keV) contours overlaid on the
optical R band images of the $z$=0.158 galaxy (P3) and the $z$=0.016
starburst galaxy (LAR5) obtained with EFOSC2.

Galaxy P3 is bulge dominated, the size of the bulge being 2--3
arcsec. Also galaxy LAR5 has a bright bulge, of similar size.  These
sizes are similar to or smaller than the Chandra PSF at the offaxis angle
where the sources were detected in ACIS-I (2.6 arcmin and 6 arcmin
respectively). Both galaxies are extended (up to 10--20 arcsec).
If the X-ray emission were connected to the outer parts of the galaxy, 
it would have been resolved by {\it Chandra}.  
Both X-ray sources are centered
on the galaxy nuclei.  The X-ray source in LAR5 appears
slightly elongated in a direction perpendicular to the galaxy major
axis, as seen in several nearby starburst galaxies (e.g. Fabbiano et
al.  1992, Dahlem et al. 1998). However, this elongation may also be due to
the degradation of the {\it Chandra} PSF at the offaxis-angle of this
source (6 arcmin).

\begin{figure*}
\centerline{
\hbox{
\epsfig{file=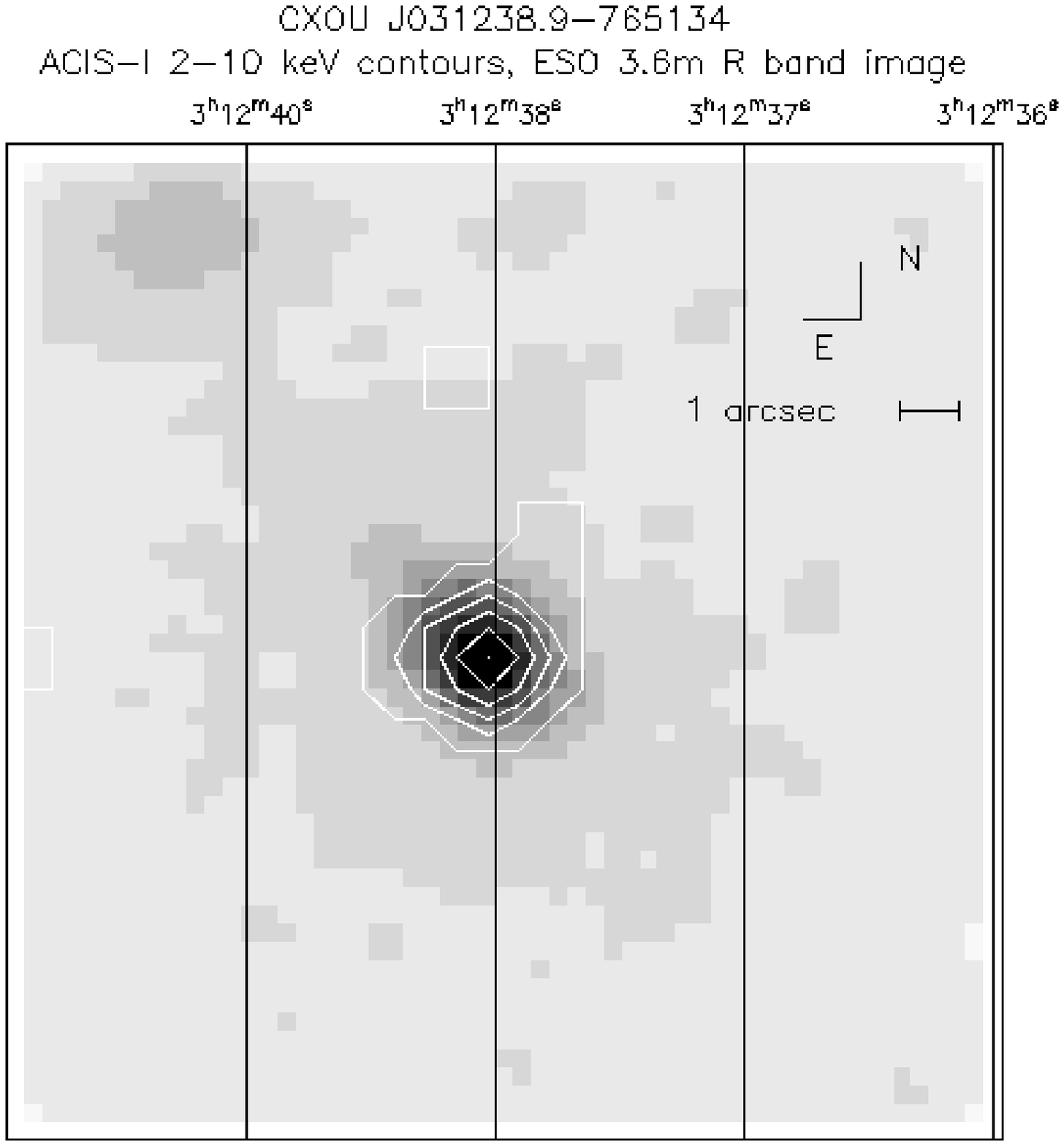, height=7.5cm,width=7.5cm, angle=0 }
\epsfig{file=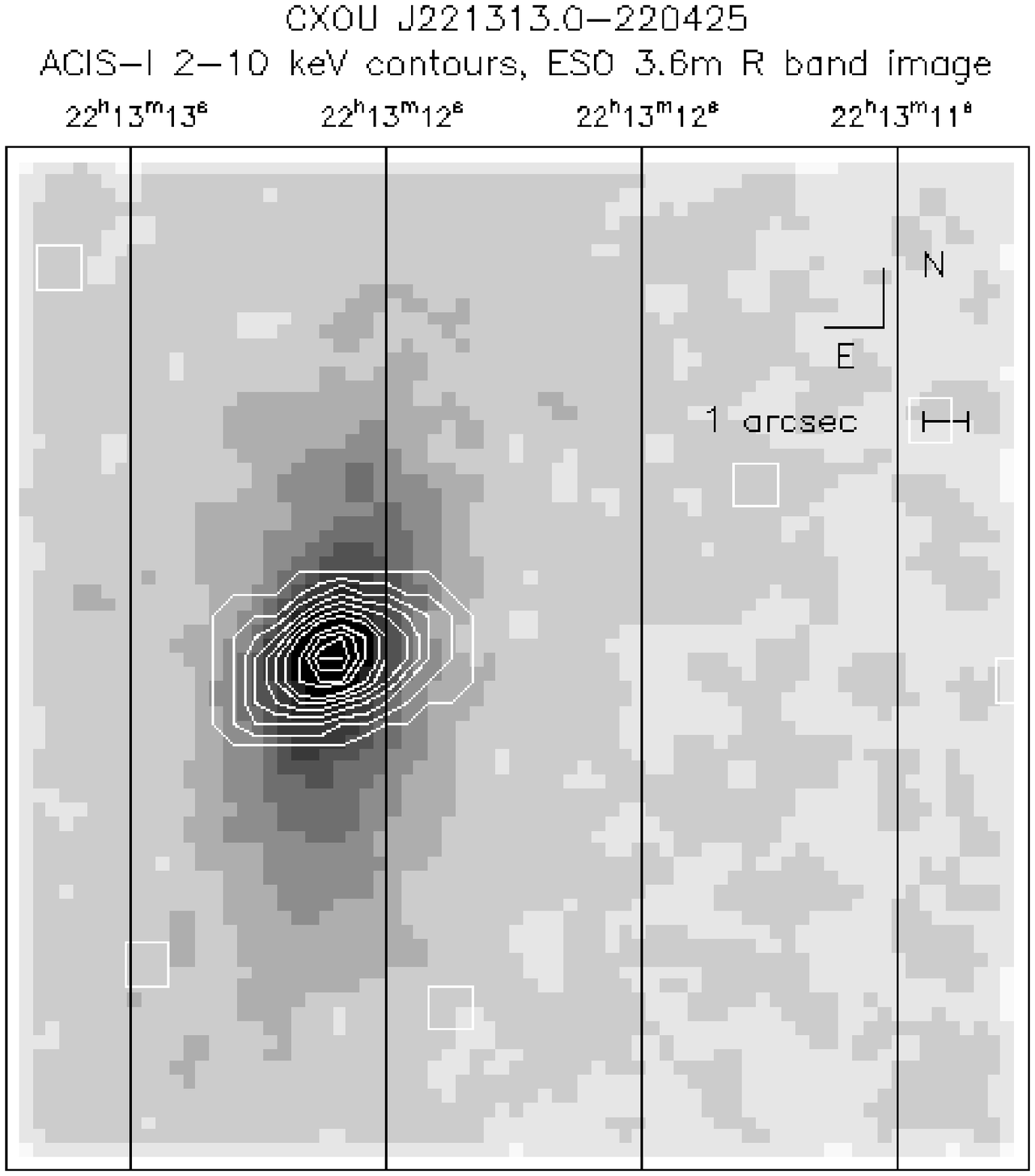, height=7.5cm,width=7.5cm, angle=0 }
}
}
\caption{
{\it Chandra} ACIS-I (contours) and R band images (grayscale) of 
CXOUJ031238.9--7651, P3 (a), and CXOUJ221313.0--220425, LAR5 (b).
The second source was observed by ACIS-I at an
off-axis angle of 6 arcmin leading to contours slightly
elongated in the East-West direction.
}
\end{figure*}

\subsection{X-ray to optical flux ratio}

Figure 3 shows the X-ray to optical flux ratios of the thirteen {\it
Chandra} sources as a function of the 2--10 keV flux, and compares them
with those of local AGN and of the AGN found in the BeppoSAX HELLAS
survey. This Figure shows that the X-ray to optical ratio of most {\it
Chandra} sources is similar to that of local and HELLAS AGN.  Their
optical magnitude is bright enough to allow redshift determination
with 4m or 8m class telescopes.

At fainter fluxes Figure 3 shows the sources recently optically
identified by Mushotzky et al. (2000). R band fluxes have been
computed assuming R-I=0.3.  The flux ratio of about 60\% of the
Mushotzky et al. {\it Chandra} sources is within the range of values
covered by brighter AGN (i.e. log$(f_X/f_R)$ from $-$2 to 1).  The
remaining 40 \% show an X-ray to optical ratio higher than that of the
brighter {\it Chandra} sources presented here and of the local AGN
(see Figure 3).  As discussed by Mushotzky et al. (2000) the nature of
these optically faint sources is mysterious. Many of them have optical
magnitude fainter than R$\sim$24 which make difficult to obtain a
precise redshift through optical spectroscopy.

The $z$=0.158 normal galaxy has an X-ray to optical ratio of 0.36, much
higher than the typical value of nearby normal galaxies (see section
5.2).

\subsection{X-ray spectral properties}

For most of the {\it Chandra} sources the total number of detected counts
is $<$100, preventing the use of proper spectral fitting
procedures to study their spectrum.  The broad band X-ray spectral
properties can only be investigated using count ratios.  If the
spectrum is parameterized as an absorbed power law and the redshift of
the source absorber is known, the column density can be
evaluated.  Following Fiore et al. (1998), we assume here that the
X-ray absorber redshift coincides with the optical 
redshift.  Figure 4 shows the softness ratio (S$-$H)/(S+H) (S=0.5--2 keV
band, H=2--10 keV band) of the ten spectroscopically identified {\it
Chandra} sources as a function of the redshift.  Errors include  counting
statistics only. Dashed lines show the
expectation of power law models with $\alpha_E=0.8$ absorbed by
varying column densities {\it in the source frame}. Two of the {\it
Chandra} sources (P3 and LAR6) have a softness ratio inconsistent with
that expected by an intrinsically unobscured power law at better than
the 90\% confidence level. In particular, 
LAR6 is likely to be obscured by a column
of $5-10\times10^{22}$ cm$^{-2}$.

It is worth noting that the column densities implied by Figure 4 are
probably lower limits to the
columns toward the nuclear hard X-ray source, because highly obscured
AGN usually have strong soft components (e.g. Schachter et al. 1998,
Maiolino et al.  1998, Della Ceca et al. 1999, Fiore et al. 2000a,b).
To quantify this possible underestimate we simulated ACIS-I
observations of three highly obscured type 2 AGN 
(NGC1068, $N_H>10^{25}$ cm$^{-2}$; the
Circinus Galaxy $N_H=4.3\times10^{24}$ cm$^{-2}$; and NGC6240, 
$N_H=2.2\times10^{24}$ cm$^{-2}$), based on ASCA and BeppoSAX results
(Matt et al.  1997, 1999, Iwasawa \& Comastri 1998, Vignati et al.
1999). The (S$-$H)/(S+H) observed by ACIS-I would have been 0.88, $-$0.08
and 0.57 respectively. The results are strongly dependent from the
sources spectral shape and redshift. At $z$=1 the simulated hardness ratios
are 0.53, $-$0.30 and $-$0.05 respectively.

We also note that relatively low values of (S$-$H)/(S+H) can be produced
by intrinsically flat and unobscured power laws 
(for example the P3 hardness ratio of 0.16 would correspond to $\alpha_E=0.2$).
Efficient X-ray follow-up of relatively bright X--ray sources
with XMM  will provide a relatively
accurate measure of the intrinsic spectrum through proper spectral fitting
allowing to remove the ambiguity between an intrinsically flat
spectrum and a spectrum flattened by absorption.

\begin{figure}
\centerline{
\hbox{
\epsfig{file=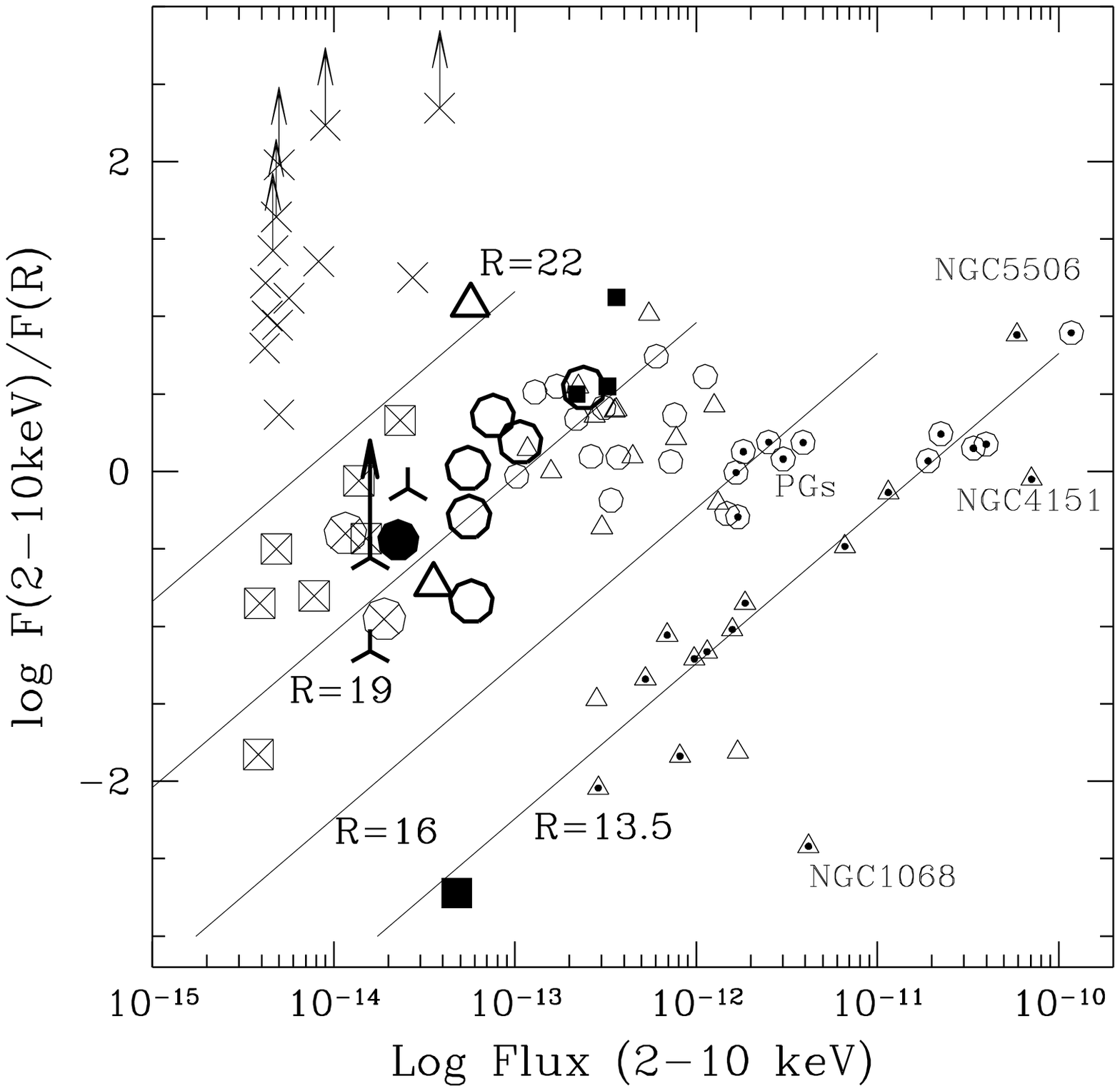, height=7.5cm,width=7.5cm, angle=0 }
\epsfig{file=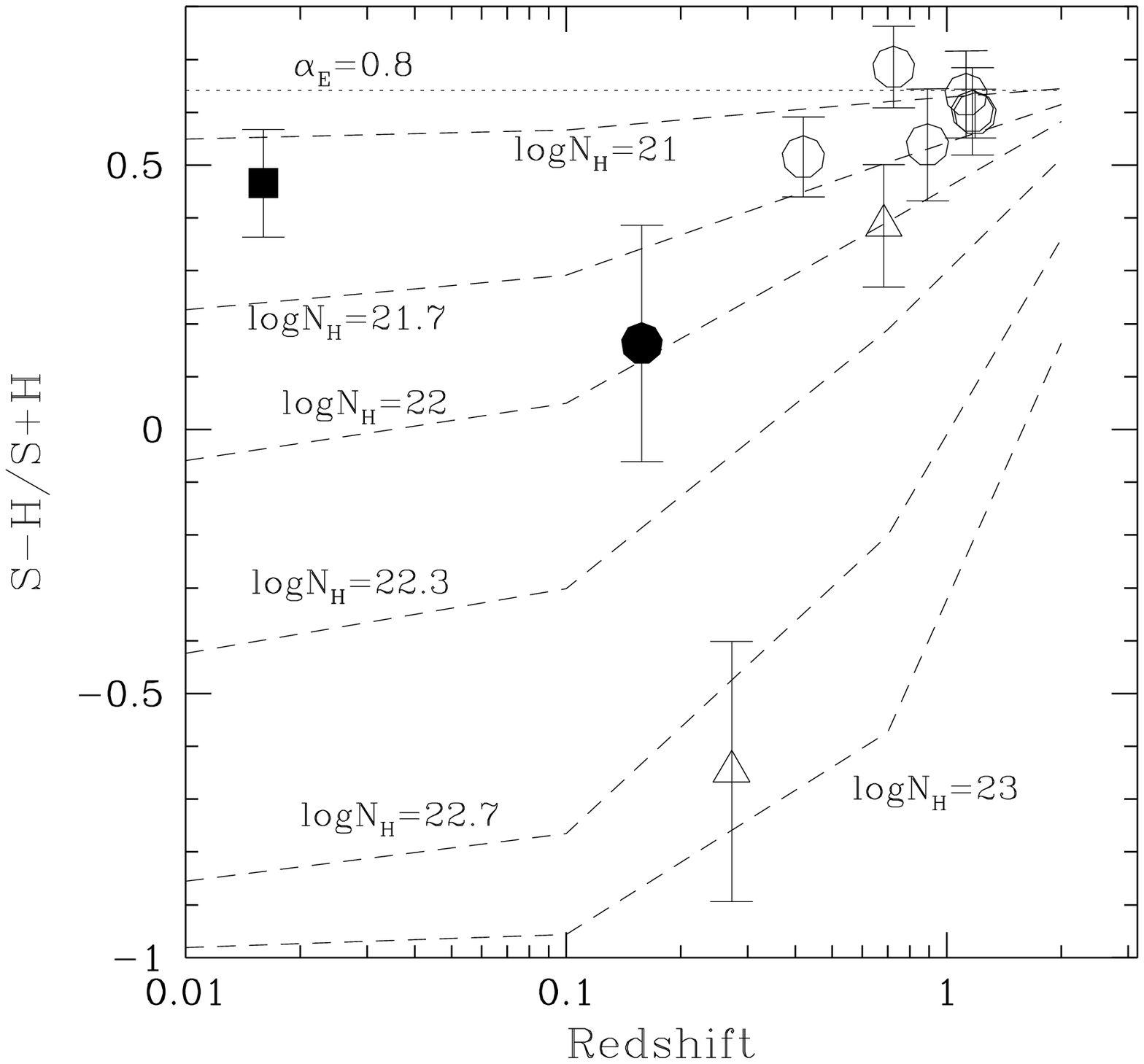, height=7.5cm,width=7.5cm, angle=0 }
}
}
\caption{
X-ray to optical ratio versus the 2--10 keV flux.  Diagonal lines
identify loci of constant apparent R magnitude.  Chandra new
identifications: big symbols; HELLAS AGN: small symbols; local 
AGN (X-ray and optically selected): dotted small symbols.
Different symbols mark identified sources: open circles = broad line
quasars and Sy1; open triangles= type 1.8-1.9-2.0 AGN; filled squares=
starburst galaxies and LINERS; filled circles= normal
galaxies. Skeletal triangles identify {\it Chandra} sources from this
paper without redshifts and optical classification.  Crosses refer to
{\it Chandra} sources identified by Mushotzky et al. (2000).
Circled Crosses = quasars; squared crosses = sources with a redshift
but unpublished optical classification.
}
\caption{
The softness ratio (S$-$H)/(S+H) as a function of redshift for
the identified sources.
Symbols as in Figure 3.
 Dotted lines show the expected softness
ratio for a power law model with $\alpha_E=0.8$.  Dashed lines show
the expectation of absorbed power law models (with $\alpha_E=0.8$ and
log$N_H$=23, 22.7, 22.3, 22, 21.7 and 21, from bottom to top) with the
absorber at the source redshift.
}

\end{figure}

\bigskip

\section{Discussion}

\subsection{Faint X-ray sources and the XRB}

One of the main goals of hard X-ray surveys is to investigate the origin
of the hard X-ray cosmic background.  The most popular model explains
the XRB in terms of a mixture of obscured and unobscured AGN (Setti \&
Woltjer 1989) following a strong cosmological evolution.

The relatively small number of sources in the present survey (only
thirteen), implies a $\sim 30\%$ statistical error on their number
density. This limits the strength of the constraints we are able to
put on AGN synthesis models for the XRB. 
However we believe it is
useful to begin addressing this issue here.

The main problem when comparing model predictions and observations is
that the flux limit of {\it count rate} limited surveys depends on the
actual intrinsic spectrum of the sources (Zamorani et al. 1988).  
Harder sources would
generally produce less counts (because of the decrease of the
effective area toward the higher energies), and therefore they would
pass a detection threshold only if their flux is higher than that of
softer sources with similar count rate.  In order to quantify this
effect we have folded a heavily absorbed (10$^{23}$ cm$^{-2}$) power law
spectrum ($\alpha_E$ = 0.8) with the {\it Chandra} sensitivity, and 
computed the flux
limit corresponding to the count rate threshold.  This turns out to be
a factor 5--6 higher than that of an unabsorbed power law.  Taking
into account the column density distribution predicted by the Comastri
et al. (1995) synthesis model and the {\it Chandra} sensitivity, the
predicted fraction of obscured (log$N_H >$ 22) sources is 20--30 \%.
The softness ratio results (Figure~4) suggest that at least 3 out of 10
optically identified sources may be absorbed by column densities equal
or higher than $10^{22}$ cm$^{-2}$. Furthermore, one of the three
unidentified objects has a hard spectrum, as can be judged from the
hard to soft X--ray flux ratio of Table 2.  
We therefore conclude that the present observations are consistent
 with the predictions of AGN synthesis models for the XRB.

\subsection {P3: an X-ray loud normal galaxy?}

A surprising result from our pilot study is the detection of a
luminous X-ray source in an otherwise normal galaxy (P3).  The X-ray
luminosities of $L_{2-10keV}\sim 3\times10^{42}$ erg s$^{-1}$,
$L_{0.5-2keV}\sim0.8\times10^{42}$ erg s$^{-1}$
are about a factor 30 higher than those expected on the basis of the
optical luminosity ($L_B=10^{10} L_{\odot}$) for both 
spiral (Fabbiano et al. 1992) and Elliptical/S0  
(Eskridge, Fabbiano \& Kim 1995; Pellegrini 1999) galaxies.

A few optically ``dull'' galaxies with strong X-ray emission have been
reported in the past (e.g. 3C264 and NGC4156, Elvis et al. 1981; J2310$-$437,
Tananbaum et al. 1997; Griffiths et al. 1996) and more recently in a 
deep {\it Chandra}
observation (Mushotzky et al. 2000). Hard power law tails have been 
also discovered in a few nearby elliptical galaxies  
(Allen, Di Matteo \& Fabian 2000).
The presence of relatively strong X--ray emission in objects with no
evidence of activity in the optical spectrum is still 
not well understood. 


One possibility would be a large contribution from a beamed
non-thermal component in the
X-ray band (Elvis et al. 1981, Tananbaum et al. 1997, Worrall et
al. 1999) as for BL Lacertae objects.
If P3 hosts a BL Lac then the non--thermal featureless 
optical continuum would reduce the height of the Calcium break
from a typical value of 0.5 found in elliptical 
and S0 galaxies to $<0.25$ (Stocke et al. 1991).
This threshold has been raised to about 0.4 by Marcha
and Browne (1995) to take into account the spread of galaxy 
luminosity and sizes. The P3 Calcium break of 0.44$\pm$0.11
does not allow to rule out the presence of a BL Lacertae object,
even if it would be a rather extreme member of its class.
If this is the case, the radio flux predicted from the 
observed X-ray flux would be in the 0.1--30 mJy range 
(Padovani \& Giommi 1996). The PMN (Griffith et al. 1991)
limit of about 50 mJy is not useful to settle this issue.

An obscured AGN could also provide a viable explanation.
Indeed the P3 hardness ratio implies a
substantial column for a typical AGN X--ray continuum.
The optical emission lines could also be completely
hidden except for a weak [OII] emission feature.
It is worth remarking that examples
of X--ray obscured AGN with neither BLR nor NLR already exist
(e.g. NGC4945, Marconi et al. 2000; NGC6240, Vignati
et al. 1999 and references therein), even if in these cases emission lines
related to starburst emission are present.

Finally the presence of a low--radiative--efficiency accretion flow 
(ADAF) might also be tenable. The putative central 
black hole mass estimated from the observed B luminosity of the 
galaxy's bulge following the Magorrian et al. (1998) relation 
is about 4$\times$10$^{8}$ M${\odot}$  
(this is actually an upper limit, as we have assumed for the bulge 
luminosity that of the entire galaxy, see Fig. 2a).  
If the X--ray emission comes from the nucleus, and it
is a sizeable fraction of the bolometric luminosity, its Eddington
ratio is of the order of 10$^{-4}$, consistent with the
ADAF regime (Narayan et al.  1998). 
An estimate of the 8.4 GHz flux has been obtained 
assuming the ADAF spectral models calculated by Di Matteo et al. 
(2000) as well as the X--ray to radio flux ratios of a few 
ADAF candidates observed at the VLA (Di Matteo et al. 1999) rescaled
to the P3 X--ray flux.
In both cases the maximum expected radio emission is 
of the order of a few mJy and thus well below the present limit.
The ADAF contribution to the optical--UV light strongly depends on 
the adopted model (see figure 2 in Di Matteo et al. 2000).
The P3 hardness ratio is consistent with a very flat power law 
($\alpha \sim$ 0.2$\pm$0.3), and a wind model is therefore to be 
preferred. If this is the case, 
the ADAF flux in the optical--UV would be much fainter 
(at least two order of magnitude) than that of the host galaxy, 
again consistent with the present findings.

\section{Conclusions}

We have carried out a pilot program to study the faint hard X-ray
source population using the revolutionary capabilities of the {\it
Chandra} satellite.  We identified ten 2--10 keV selected sources from
two 4-chip, medium deep {\it Chandra} fields covering about 0.14
deg$^2$ of sky at fluxes in the range $1.5-25\times10^{-14}$ \cgs, a
factor of 3 fainter than previous ASCA and BeppoSAX surveys.
Recently, Mushotzky et al. (2000) reported detection of faint X-ray
sources from a single 1--chip field (0.0175 deg$^2$) 
at fluxes $0.3-3\times10^{-14}$ \cgs.  Our results fill the
gap between the shallow BeppoSAX and ASCA surveys and the deep {\it
Chandra} field.

Almost all sources have an optical counterpart within 2 arcsec. Optical
spectra allow us to measure their redshifts, and assess their optical
classification.  We find six broad line quasars, two emission line AGN
(LAR6 and P4), one starburst galaxy (LAR5), and one apparently normal
galaxy at $z$=0.158 (P3).  LAR6 and P4 are likely to be obscured in
X-ray by a column densities of $\approx10^{23}$ cm$^{-2}$ and
$\approx10^{22}$ cm$^{-2}$ respectively.
The X-ray source in the $z$=0.158 normal galaxy P3 may be covered by a
column density of about $10^{22}$ cm$^{-2}$ too.

The spatial extension of all X-ray sources is smaller than a few
arcsec in all cases, and it is roughly consistent with 
the {\it Chandra} PSF. The X-ray sources associated with the
$z$=0.016 starburst galaxy LAR5 and the $z$=0.158 normal galaxy P3 appear
coincident with the galaxy nuclei.

The X-ray to optical luminosity ratio of P3 is higher by a factor of at
least 30 than those of normal galaxies, while it is similar to those
of AGN.  The high X-ray luminosity and the lack of optical emission
lines suggest an AGN in which either continuum beaming is important or
emission lines are absorbed or not efficiently produced. 
In any case, objects like P3 would be missed or ignored in optical
surveys or in X-ray surveys with large error boxes. It is only thanks
to the new revolutionary capabilities of {\it Chandra} that this kind
of sources {\it can} be detected and identified.

Based on the ASCA and BeppoSAX surveys at fluxes $>5\times10^{-14}$
\cgs and on our first {\it Chandra} identifications, 
which push the flux limit down to $\sim2\times10^{-14}$ \cgs, the hard
X-ray sky appears populated by a large fraction of broad line AGN
(about 50\%), by a mixture of intermediate AGN (type 1.8-2.0 and composite
starburst/AGN) and, most intriguingly, also by X-ray luminous 
apparently normal galaxies. These populations span ranges of X-ray and optical
properties wider than previously thought.

\bigskip

{\it Acknowledgements}

The results presented in this paper are made possible by the
successful effort of the entire {\it Chandra} team. In particular we
thank the XRT and ACIS teams for building and calibrating the high
resolution mirror and the CCD camera, the CXC team, and in particular
A. Fruscione, for the quick data reduction and archiving.  We thank
P. Giommi, S. Molendi, G. Fabbiano, E. Giallongo, M. Mignoli,
L. Stella, M. Vietri and H. Tananbaum  for useful
discussions and L.A. Antonelli for help in the ESO observation
preparation. We also thank the referee, Gianni Zamorani,  
for his detailed and constructive comments which improved 
the quality of the paper.
This research has made use of the NASA/IPAC Extragalactic
Database (NED) which is operated by the Jet Propulsion Laboratory,
California Institute of Technology, under contract with the National
Aeronautics and Space Administration.  This work is partly supported
by the Italian Space Agency, contract ARS--99--75 and by the Ministry
for University and Research (MURST) under grant
COFIN--98--02--32. M.E., F.N, and M.C. acknowledge support from NASA
contract NAS8--39073.

\end{document}